# Using Zone-Disjoint Multi-Path Routing Algorithm for Video Transmission over Ah-Hoc Networks

Ali Asghar Khavasi
Networks and Communications Laboratory
Electrical, Computer & IT Engineering Department
Zanjan Azad University
Zanjan, Iran
ali.a.khavasi@gmail.com

Nastooh Taheri Javan
Ad-Hoc Networks Laboratory
Computer Engineering Department
Amirkabir University of Technology
Tehran, Iran
nastooh@aut.ac.ir

## ABSTRACT

Finding multi-path routes in ad hoc networks due to their grid topology seems to be a trivial task, but because of CSMA/CA effects in these networks found paths are not completely disjoint unless an appropriate algorithm have taken into account. If such an algorithm provided and designed carefully it could improve multi-path video transmission over these kinds of networks. By using node-disjoint paths, it is expected that the end-to-end delay and BER in each case should be independent of each other. However, because of natural properties and medium access mechanisms in ad hoc networks the end-to-end delay and also BER between any source and destination depends on the pattern of communication in the neighborhood region. In this case some of the intermediate nodes should be silent to reverence their neighbors and this matter increases the end-to-end delay. To avoid this problem, multi-path routing algorithms can use zone-disjoint paths instead of node-disjoint paths. In this paper we demonstrated a new multi-path routing algorithm that selects zone-disjoint paths that in addition used for video multi-path transmission over ad hoc networks. It is shown that by using this new algorithm along with choosing an appropriate scheme for video transmission over discovered paths, our approach receives enhanced results comparing previously used algorithms.

## Categories and Subject Descriptors

C.2.2 **[Computer Communication Networks]**: Network Protocols- *Routing Protocols*; H.5.1 [Information Interfaces and Presentation]: Multimedia Information Systems- *Video*.

## General Terms

Algorithms

## 1. INTRODUCTION

Ad-Hoc Networks are such a kind of Wireless networks in which nodes are mobile and there is no pre-determined organizer node in their structure [1]. These networks are often created instantaneously and mobility of nodes, their entering and leaving to/from the network and therefore emersion and devastation of paths in the network, feature a semi-random state. Existing paths in these networks usually consist of several hops which each of them does the routing process along the path. Nowadays, these networks have gained wide applications in various fields. Most of these networks are applied to some cases that are not possible to set up a pre-setup network. Some examples of such type of networks are found in Military applications, Rescuing and Revealing Operations especially in unexpected events and also some applications in conferences, roads, education and entertainments. In [2], some samples of these applications are presented.

To provide end-to-end communication throughout the network, each mobile node acts as an intermediate router forwarding messages received by other nodes [1]. The wireless radio link may be interrupted owning to one of the mobile nodes moving out from the original radio radius, running out of battery or being turn off by the user. The routing path between sender and the receiver can also be fractured.

Design of efficient routing protocols is the central challenge in such dynamic wireless networks. Routing protocols for MANETs can be broadly classified into reactive (on-demand) and proactive algorithms [3]. In reactive protocols, nodes build and maintain routes as they are needed but proactive routing algorithms usually constantly update routing table among nodes. In on-demand protocols, nodes only compute routes when they are needed. Therefore, on-demand protocols are more scalable to dynamic, large networks. When a node needs a route to another node, it initiates a route discovery process to find a route.

Numerous well-studied ad hoc wireless routing protocols, such as Dynamic Source Routing (DSR) [4] or Ad hoc On-Demand Distance Vector Routing (AODV) [5], rebroadcast the "Path Discovery Messages" and seek another routing path. Newly discovered paths may become un-useful even before the start of routing if network topology changes too frequently. Moreover, the network topology may change again before the last topology updates are propagated to all intermediate nodes.

One of interesting properties of ad-hoc networks is that in the face of existing problems, they give this chance of feasibility of using mesh structure for transmission, inasmuch as it is possible to divide data stream to some sub-streams and then transmit them to various paths on the network. Works done by Wang, and others [6, 7, and 8] can be considered as the most brilliant works in this field which take advantage of video layer coding techniques, Automatic Repeat Request (ARQ) and Multi-Description Coding (MDC). However, in practice, there should be an algorithm that could be able to provide such a degree of independency between paths and consequently, this very essential hypothesis faces some challenges.

As we know there are two problems in wireless networks, known as "hidden station" and "exposed station". For solving these problems, CSMA/CA [9] protocol has been suggested. In 802.11 standard, this protocol is used for access to the channel. Due to transferring RTS[2] and CTS[3] packets between nodes in this protocol, some of the nodes don't transfer the data and as a result the delay is increased.

As an example, consider figure 1 that shows an imaginary LAN with ten nodes [11]. In this figure radio range of every node is distinguished and the dotted line shows the relation between nodes. In other words, the dotted lines between two special nodes show that they are located in each other radio range.

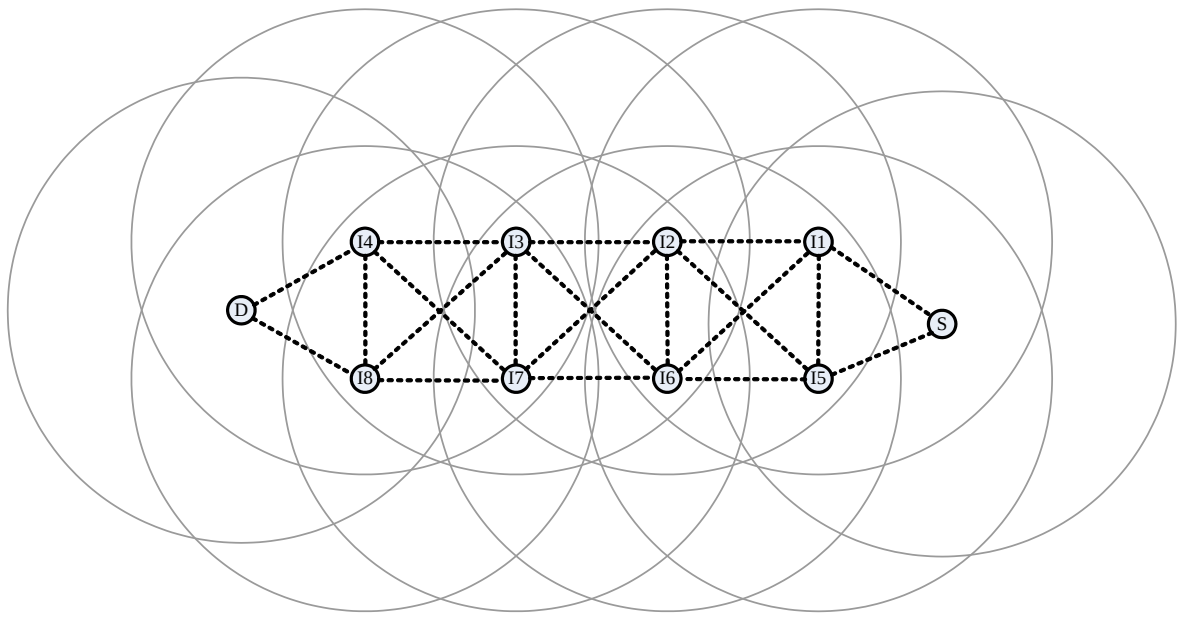

**Figure 1. Node-Disjoin paths [11].**

There are two node-disjoint paths, S-I1-I2-I3-I4-D and S-I5-I6-I7-I8-D, between D and S, which transferring the data in one path is not completely separated from the other path. In this case, the delay of every path is related to the other path traffic Because of transferring RTS[2] and CTS[3] packets between nodes of network in order to collision avoidance and solve hidden station and exposed station problems. As a result some of the station of a path in order to receive CTS from a node in opposite path should postpone their sending.

To solve this problem, we can use zone-disjoint paths instead of node-disjoint paths. Two routes with no pair of neighbor nodes are called zone-disjoint in terminology. In [12,13] the authors proposed a method for distinguishing the zone-disjoint paths, that uses directional antenna, but most of the present equipment is not equipped with directional antenna. In this paper a multi-path routing algorithm has suggested. In this approach, by using omni-directional antenna, the zone-disjoint paths are recognizable and can be used to send co-process data simultaneously.

A Multi-Path Routing Algorithm by the name of ZD-MPDSR has been offered in [14]. ZD-MPDSR discovers Zone-Disjoint paths between source and destination nodes, and source node uses these paths for simultaneous sending of date to destination. Although that algorithm gets some improvement of decreasing of end-to-end delay, but the overhead of its routing is so height. On the one hand, its route discovery process do with height delay but this delay is being compensated in data sending phase.

In this paper we discussed the route discovery process in ZD-MPDSR which caused the delay of route discovery to be decreased. In addition, a new scheme for video transmission over these routes demonstrated that exploits benefits of these heuristic path selections to improve transmitted video quality.

The rest of this paper is organized as follows. The following section deals with the related works. Section III describes the proposed protocol and transmission mechanism in detail. Performance evaluation by simulation is presented in section IV and concluding remarks are made in section V.

## 2. RELATED WORKS

In 1989, Ghanbari [15] showed that video stream can be coded to two (or multiple) separate streams inasmuch as that one of streams has a higher priority from a video point of view. In other words, it is possible to consider vital video information as a base layer that receiver would be able to build a generic image of video and then other layers would be assumed as enhanced layers that carry more information related to images details. Afterwards, to show if we could prioritize these layers based on significance in the network (for example in queues while congestion is emerged), we will be able to reach a technique for resilience in the face of error.

In presented methods by Wang and others, in the first method which is called "selection of reference image based on feedback", a generic encoder is used for creation of stream. This encoder is aware of network status and based on this information, its own policies to output stream with RPS method is exerted. Here, sender separates even and odd frames from output stream and sends even frames through a path that we call them "even path" here, and sends odd frames through an odd path to the receiver in the same way. The encoder checks two paths status permanently, in such a way that upon reception of the first NACK from any possible paths, that path is marked under the label of "bad" and thus while encoding, those frames are used as a reference frame which has been sent on the "good" path- path whereby has received ACK from recently, or ACK related to those frames was received in any possible way. A "bad" path will be re-formed to a "good" path only if an "ACK" is received in that path and "good" path by receiving a "NACK" will change to "bad" status. RPS scheme presents a proper equilibrium between encoding efficiency and error-resilience. Certainly, mentioned scheme is only applicable in online encoding status, because in this scheme, encoder should permanently be aware of the network status and it should change the encoding technique if needed. In figure 2, performance process of this method has been presented schematically.

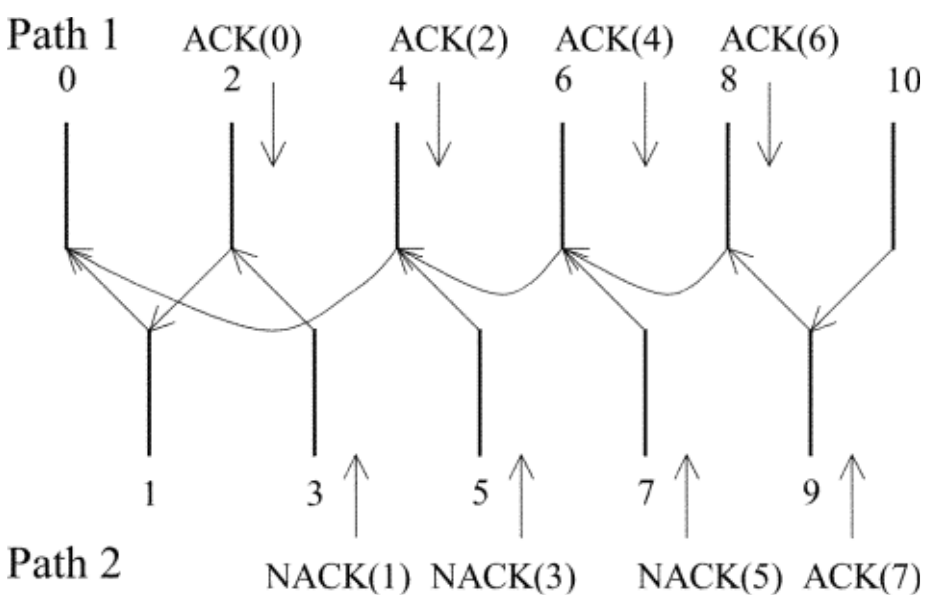

**Figure 2. Illustration of the RPS scheme. The arrow associated with each frame indicates the reference used in coding that frame [7]**

Second method, is called "Layer encoding by selective ARQ". In this method, encoder encodes video to two or multiple streams. In this case, one of streams contains vital and basic information for video demonstration and thereby it is called "Base Layer" and other layers transmit information relevant to details that reception of them by receiver upgrades video quality; therefore they are called "enhanced layers". In the proposed scheme, the assumption is based on the fact that there are only 2 paths between sender and receiver. However, this scheme performs for more paths in the same way. Here, packets related to base layers are sent through a better path that has a lower packet loss rate and packets relevant to enhanced layer are sent through other layer. In the case of base layer packets loss, receiver sends an ARQ message to sender in order to have sender re-transmit those packets. Here, lost packets are sent through enhanced layer path to the receiver. This leads to (1) Transmission bit range does not change through base layer path and only bit transmission rate of enhance layer decreases which is called LC (Layered Coding) with ARQ; (2) That in the case that all packets are lost in base layer path due to congestion of a middle node, re-dispatch of packets in this path will cause continuation of congestion and so packets drop continues; (3) Third case is mentioning to independency of two paths which causes possibility of error occurrence in both paths that decreases simultaneously and hence, dispatch of lost packets of base layer path on enhanced layer path, will decrease the probability of their loss. The advantage of this method is that there is no need to change in encoding policies upon path conditions change and consequently, complexity of encoder decreases.

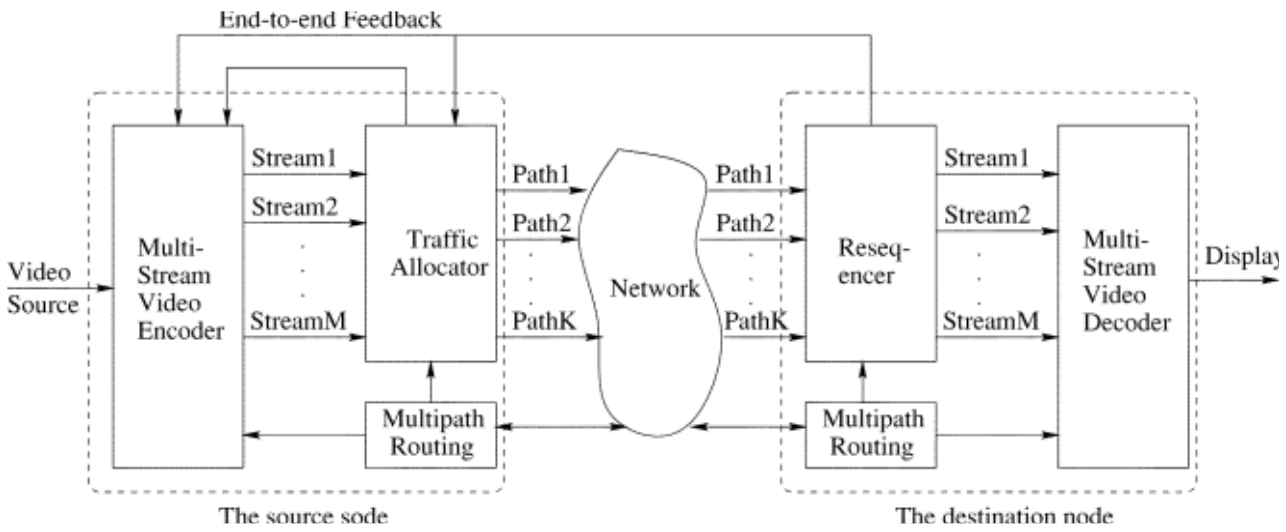

**Figure 3. General Structure for MD Coding with Multi-Path Transmission [6]**

However, this method is appropriate for video streaming state in which video stream is encoded in an offline manner and there is only a need to a video streaming system. In figure 3, generic structure of this method is presented.

In the third method that is called "multi-description coding", technique of MDC is exerted. Here, encoder produces two descriptions. This work is done by dispatch of even frames as a description and odd frames as another description. Assume that while we are encoding image n, encoder makes two types of prediction:

• Prediction produced by Linear Superposition of two previous encoded frames- n-1 and n-2 -which is called "Central Prediction".

• Prediction produced by previous encoded image which was put in this description- image n-2 -which is called "Side Prediction".

Then, encoder codes two signals for image n. These signals are central prediction error (difference between image n and central prediction) and reference mismatch signal (essential difference between central and side predictions). First description includes central prediction errors and reference mismatch signals for even frames and second description includes all the same for odd frames. When two descriptions are received, the decoder can produce central prediction again and re-create image with adding central prediction error to central prediction. When only one of descriptions reaches to the receiver, the decoder can only produce side description and image is decoded with both signals of central prediction error and reference mismatch. Redundancy of this encoder can be set by prediction coefficients used for central predictions and multiplexer used for reference mismatch.

In comparison with Multiple Description Motion Compensation (MDMC), there is no need to channel/network encoder for providing various protective layers and feedback as well. Acceptable quality can be obtained even when frequent packet loss occurs until this packet loss does not occur in both paths simultaneously and sufficient amount of redundancy with appropriate selection of prediction coefficients and signal multiplexer of mismatch signal is added.

In [16], it has shown that routing of multiple paths may not be the best way for video transmission of today and future ad-hoc networks while these networks are functioning based on IEEE 802.11, which is used now. Indeed, in [15] it has been demonstrated that routing of multiple paths, except in some especial scenarios, performs worse than single-path along with single-description. It is because of the fact that in those used multipath algorithms paths are not disjoint and so they has effects on each other, though by using a comprehensive disjoint multipath routing algorithm it could be eliminated.

The goal of SMR [6] is the finding maximally disjoint multiple paths. SMR is an on-demand multi-path source routing that is similar to DSR [4]. To discovery the routing paths, the source, at first, broadcasts the RREQ to every neighbor. When the RREQ is delivered to a node, the intermediate node's ID is included into packet. Then node, receiving RREQ, re-broadcasts it to every outgoing path. In this algorithm, the destination sends a RREP for the first RREQ it receives, which represents the shortest delay path. The destination then waits to receive more RREQs. From the received RREQs, the path that is maximally disjoint from the shortest path is selected and destination sends a RREP fore the selected RREQ. In SMR the intermediate nodes do not reply to RREQs, this is to allow the destination to receive all the routes so that it can select the maximally disjoint paths.

Our proposed and utilized algorithm is an improvement of ZD-MPDSR [11]. ZD-MPDSR that is based on DSR uses several Zone-Disjoint paths between source and destination to send data over multiple paths simultaneously. Like ZD-MPDSR, intermediate nodes do not need any Route Cache Table. The RREP packet in both ZD_MPDSR and IZM-DSR need the ActiveNeighborCount field to find zone-disjoint paths. Like ZD_MPDSR, in IZM-DSR intermediate nodes need RREQ_Seen table with a few changes.

In following section this algorithm along with a scheme for video transmission over disjoint multi-paths are proposed.

# 3. PROPOSED SCHEME

## 3.1 IZM-DSR Algorithm

In on-demand algorithms, when source has data to send, but no route to the destination is known, source sends a route request packet to all its neighbors.

In IZM-DSR, each intermediate nodes that receive RREQ, insert this packet in RREQ_Seen Table and increase the Counter field in RREQ_Seend Table.

When destination receive RREQs, it creates a RREP with the ActiveNeighborCount value sets to zero, and send it to the source along the reverse path that included in RREQ.

Each intermediate node that receives a RREP packet uses the count field in RREQ_Seen Table to update the ActiveNeighborCount field.

When a RREP receives to the source, the source waits for a certain time to receive all other RREPs. After that source can select some paths with less ActiveNeighborCount field from received RREP and can send data over selected routes.

To understand the operation of nodes in this Algorithm, pseudo code of source node, intermediate nodes and destination node are shown in figure 4,5 and 6 respectively.

1. When no route information to the destination is known send RREQ.
2. Wait for return RREP packets from destination.
3. If return first RREP from destination with adjust a timer, wait specific duration for receiving rest of RREP packets.
4. After expired time of timer, ascending sort the received RREP packets with ActiveNeighborCount field exist on it.
5. According the consideration contract in Load Balancing field, to needed numbers, select path from first of sorting list of RREP packets.
6. Begin sending data to destination as simultaneous, from selected paths in step 5.

**Figure 4. Pseudo code for the source node in IZM-DSR.**

1. After receipt RREQ packets, according its Routing Algorithm, instead of each RREQ packet, send a RREP packet with primary amount of ActiveNeighborCount equal Zero to source.

**Figure 5. Pseudo code for the intermediate node in IZM-DSR**

1. If receive a RREQ packet, and this packet is new and acceptable packet, insert characteristics of this RREQ packet in RREQ_Seen table and equal Zero Counter field of similar of it.
2. If receive a repetitive RREQ packet, increase one unit, Counter field regarding to this RREQ in RREQ_Seen table.
3. According politic of Route Discovery, if RREQ packet is able to re-broadcast, broadcast it for all.

**Figure 6. Pseudo code for the destination node in IZM-DSR**

## 3.2 Video Transmission Scheme

We used layered video coding with priority exploiting simplest method, data partitioning. Using this method decreases process overhead of encoder while it is convenient for offline video streaming that is the case of our scheme. Video frames are divided into two categories: Base Layers (Frames I and P) and Enhanced Layers (B Frames). As illustrated, IZM-DSR finds disjoint paths with variant ActiveNeighborCounts that has direct effect on end-to-end delay and packet-loss probability. Further, these parameters are important for video delivery when we want to choose a path, for example, for a base layer to deliver its packets/frames on it.

In our scheme we choose two paths with low ActiveNeighborCount (low loss probability) for base layer that consists of I and P frames and other disjoint paths for enhance layer that consists of B frames. In addition, receiver sends an ARQ message when an I frame is lost. In such a situation, source sends that base layer's I frame in all available disjoint paths to decrease the loss probability of these important frames.

Packet losses occurs frequently in situations like topology change (path breaks) and/or congestion in intermediate nodes of the path, so it usually takes a while to get back to the previously normal state. Therefore, in our scheme source continues to send all of the I frames in all paths for a threshold equals by playback buffer size (1000 ms in our scheme). If I frame loss continues after a threshold, the source sends frames for another time slot while the condition becomes false (no I frame loss).

# 4. PERFORMANCE EVALUATION

In order to demonstrate the effectiveness of our algorithm, we evaluate our proposed scheme and compare its performance to SMR[10] and normal DSR[4].

## 4.1 Simulation Environment

Our simulation runs on the GloMoSim simulation platform [17] and evalvid [18] framework for video quality evaluations. The GloMoSim library is a detailed simulation environment for wireless network systems.

Our simulation environment consists of N mobile nodes in a rectangular region of size 1000 meters by 1000 meters. The nodes are randomly placed in the region and each of them has a radio propagation range of 250 meters.

The radio model to transmit and receive packets is RADIO-ACCNOISE which is the standard radio model used. The IEEE 802.11 was used as the medium access control protocol.

The random waypoint model was adopted as the mobility model. In the random waypoint model, a node randomly selects a destination from the physical terrain. It moves in the direction of the destination in a speed uniformly chosen between a minimum and maximum speed specified. After it reaches its destination, the node stays there for a time period specified as the pause time. In our simulation, minimum speed was set constant to zero.

For video evaluation, Foreman video was chosen as the video sequence sample. We used QCIF sized version with 10fps and ffmpeg [19] video codec was used as a compression software tool along with evalvid general modules. The buffer size of decoder was set to 1000 ms that is reasonable size for offline video streaming.

## 4.2 Simulation Results

We compared our scheme with SMR and DSR protocols. Figures 7 and 8 show the results of these comparisons in packet delivery ratio and end-to-end delay in average respectively, under different node speeds.

In another point of view, for video evaluation we chose the scenario with maximum speed of 20 m/s and applied our video transmission scheme to it. Figures 9 and 10 show the five top paths chosen from all found paths considering five lowest ActiveNeighberCounts. Rods with lowest ActiveNeighberCounts among these five are chosen for base layer transmission and the others for enhancement layer and/or occasionally base layer packets (described in previous section) by the scheme. As illustrated in figures 9 and 10, base layer packets (I and P) tolerate low average delays and loss ratio compared to three other rods with 12, 12 and 15 ActiveNeighberCounts respectively.

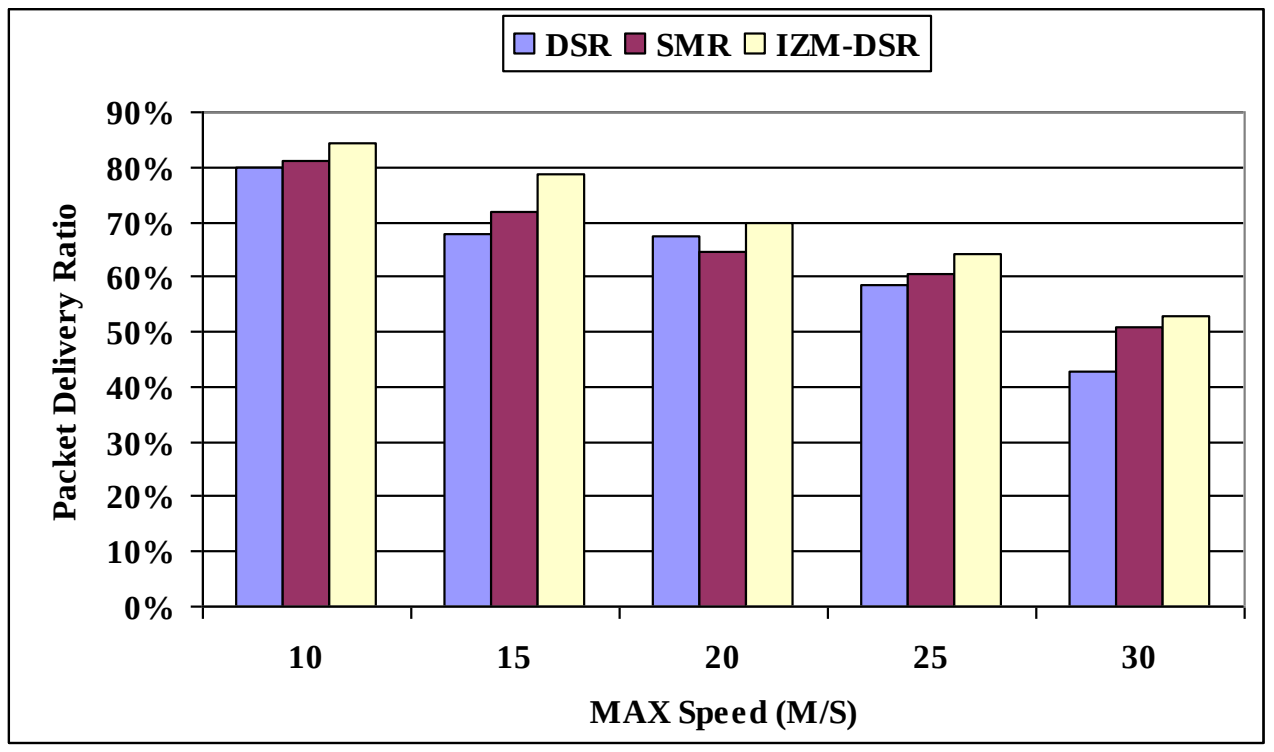

**Figure 7. Packet Delivery Ratio vs. Max Speed**

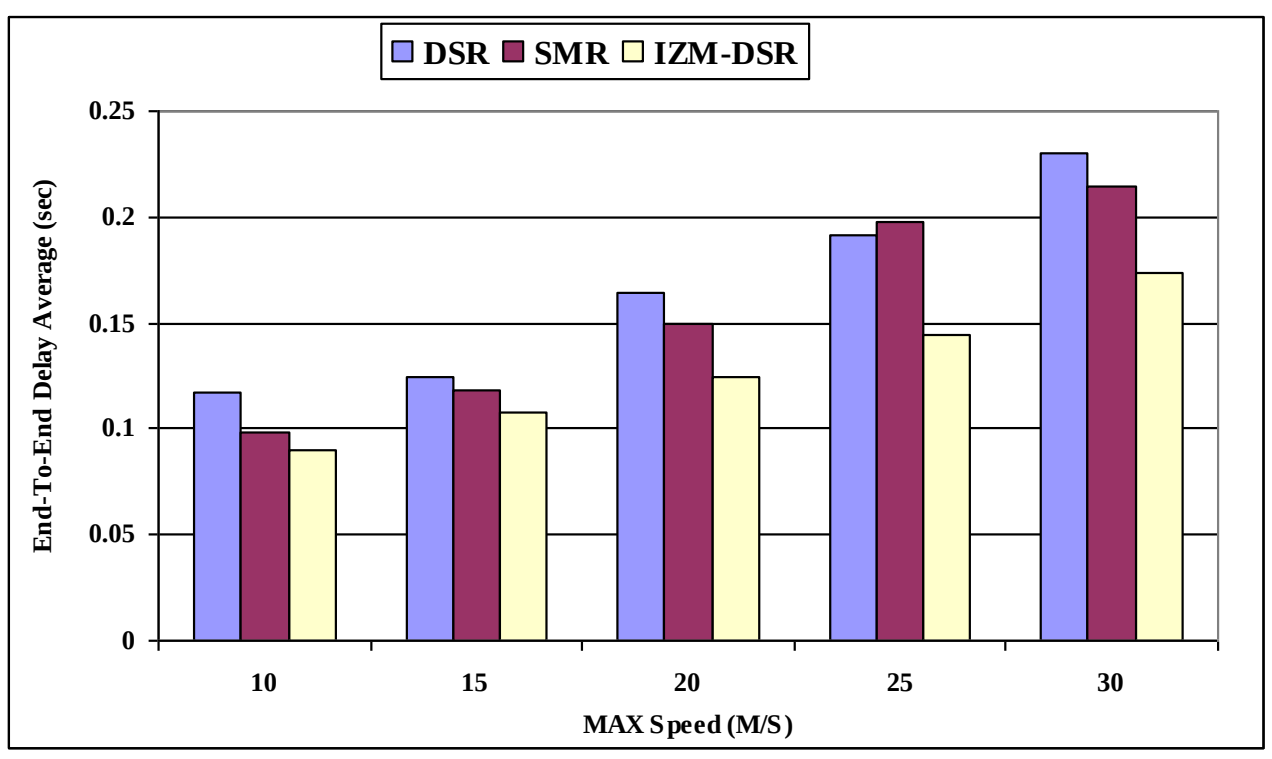

**Figure 8. End-to-End Delay Average vs. Max Speed**

Also we computed PSNR metric for proposed scheme in this specific scenario and compared it with results form SMR and DSR protocols. Table 1 shows this comparison in terms of PSNR and MOS metrics.

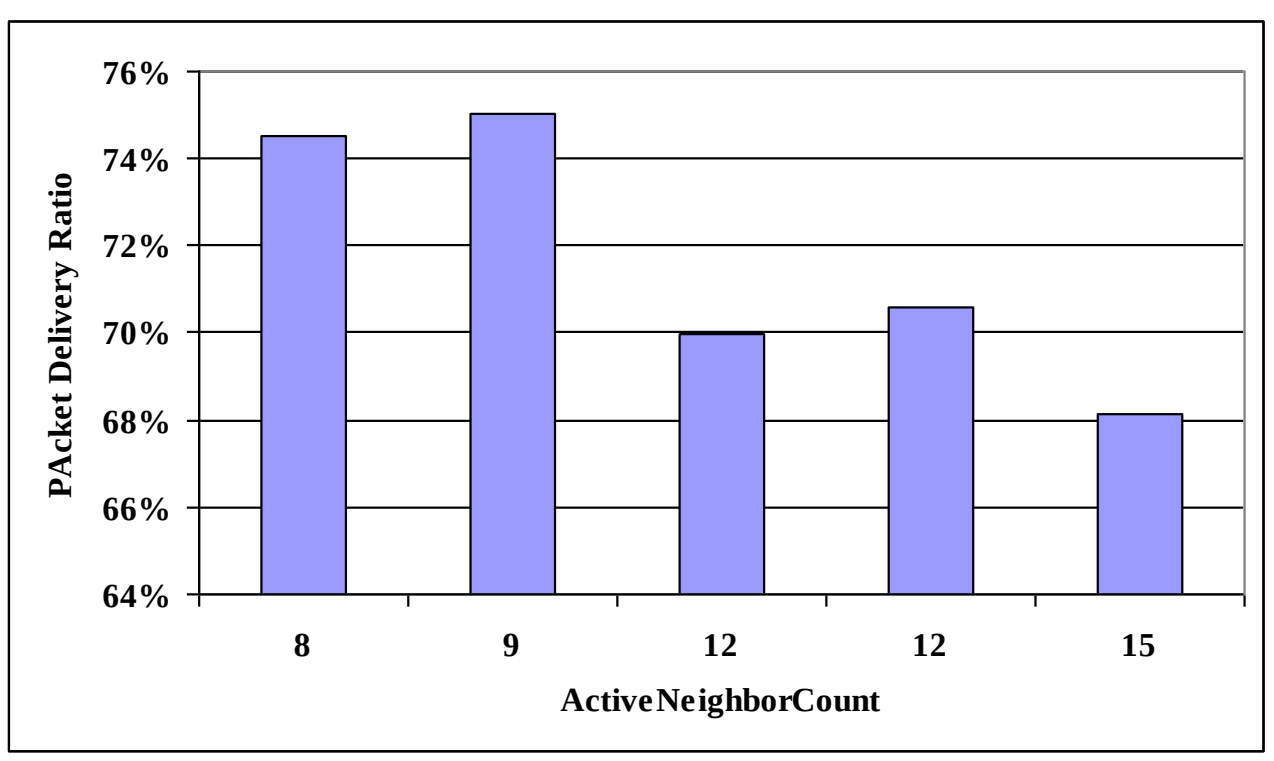

**Figure 9. Packet Delivery Ratio for five smallest ActiveNeighborCount field in IZM-DSR**

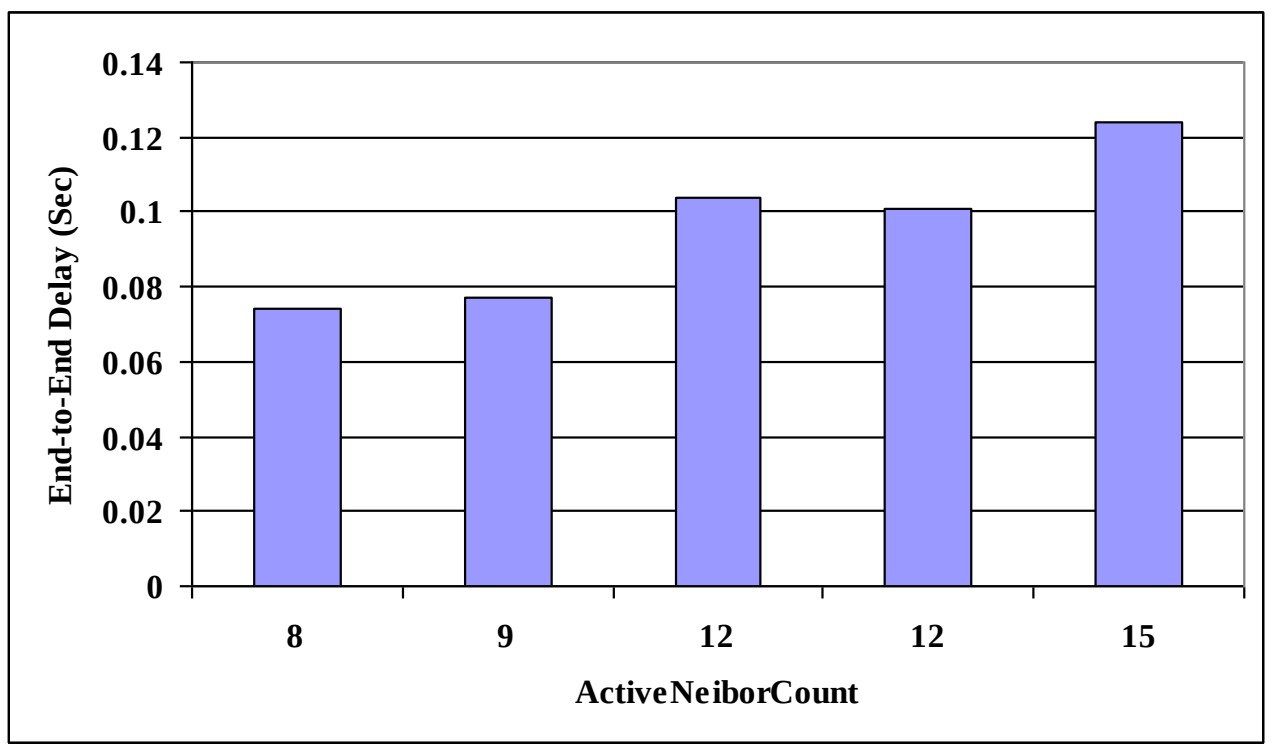

**Figure 10. End-to-End Delay for five smallest ActiveNeighborCount field in IZM-DSR**

**Table 1. PSNR Results**

| Metric / Protocol | PSNR | MOS |
|---|---|---|
| IZM-DSR | 31.4 dB | Good (Near Border Line) |
| SMR | 30.5 dB | Poor (Near Border Line) |
| DSR | 29.8 dB | Poor |

# 5. CONCLUSION

In this paper we presented a new scheme for video transmission over ad hoc networks using IZM-DSR disjoint multipath routing protocol. Video transmission can take advantages of multipath transmission and combine it with layered coding methods in which important video data, like base layer, could be transmitted from the best path among the other and less important data, enhancement layer, has been sent over other available paths. The main point is the fact that these paths should be disjoint logically so that they have no effect on and don't interfere with each other. Disjoint multipath routing algorithms try to find these paths and give a designer of an application some specific metrics to choose between those found paths. As shown in this paper IZM-DSR works perfectly in this situation by providing almost disjoint paths that in addition can be used for video delivery based on layered mechanism over mobile ad hoc networks.

# 6. REFERENCES

[1] S. Sesay, Z. Yang, J. He. "A Survey on Mobile Ad Hoc Wireless Network," Information Technology Journal, vol. 2, pp. 168-175, 2004.

[2] S. Obana, B. Komiyama, K. Mase, “Test-Bed Based Research on Ad Hoc Networks in Japan,” IEICE Trans. Commun, vol. E88-B, No. 9, September 2005.

[3] E. Royer, C. Toh, "A Review of Current Routing Protocols for Ad-hoc Mobile Wireless Networks," IEEE Personal Communication Magazine, pp. 46-55, 1999.

[4] C. E. Perkins, E. M. Belding-Royer, and S. Das, "Ad hoc On-Demand Distance Vector (AODV) Routing," RFC 3561, July 2003.

[5] B. Johnson, D. A. Maltz. "Dynamic Source Routing in Ad-Hoc Wireless Networks," Mobile Computing, vol.353, pp. 153-81, 1996.

[6] S. Mao, S. Lin, S. S. Panwar, Y. Wang, and E. Celebi, “Video Transport Over Ad Hoc Networks: Multistream Coding Multipath Transport,” IEEE Journal on Selected Areas in Communications, vol. 21, no.10, 2003.

[7] S. Mao, S. Lin, S. S. Panwar, Y. Wang, and Y. Li, “Multipath Video Transport Over Ad Hoc Networks,” IEEE Journal on Wireless Communications, 2005.

[8] S. Mao, Y. Wang, S. Lin, S. S. Panwar, “Mobicom Poster: Video Transport over Ad Hoc Networks with Path Diversity,” Mobile Computing and Communication Review, vol.7, n.1.

[9] Colvin, "CSMA with Collision Avoidance," Computer Communication, Vol. 6, pp. 227-235, 1983.

[10] S. j. lee, M. Gerla, "Split Multipath Routing with Maximally Disjoint Paths in Ad hoc Networks," in proceedings of IEEE International Conference on Communication (ICC), pp. 3201-3205, Helsinki, Finland, 2001.

[11] N. Taheri Javan, R. KiaeeFar, B. Hakhamaneshi, M. Dehghan, “ZD-AOMDV: A New Routing Algorithm for Mobile Ad hoc Networks,” in Proceedings of 8th IEEE/ICIS 2009 (International Conference on Computer and Information Science), pp. 852-857, Shanghai, China, 2009.

[12] S. Roy, D. Saha, S. Bandyopadhyay, Tetsuro Ueda, S. Tanaka, "Improving End-to-End Delay through Load Balancing with Multipath Routing in Ad Hoc Wireless Networks using Directional Antenna," in proceedings of IWDC 2003: 5th International Workshop, LNCS, pp. 225-234, 2003.

[13] S. Bandyopadhyay, S. Roy, T. Ueda, k. hasuike, "Multipath Routing in Ad hoc Wireless Networks with Directional Antenna," Personal Wireless Communication, vol. 234, pp. 45-52, 2002.

[14] N. Taheri Javan, M. Dehghan, "Reducing End-to-End Delay in Multi-path Routing Algorithms for Mobile Ad Hoc Networks," in proceedings of MSN 2007: Third International Conference on Mobile Ad-hoc and Sensor Networks, LNCS 4864, pp. 715-724, Beijing, China, 2007.

[15] M. Ghanbari, “Two-Layer Coding of Video Signals for VBR Networks,” IEEE Journal on Select. Areas Communications, vol. 7, pp. 801–806, June 1989.

[16] I. F. Diaz, D. Epema, J. D. Jough, “Multipath Routing and Multiple Description Coding in Ad-Hoc Networks: A Simulation Study,” ACM 1-58113-959-4/04/0010, 2004.

[17] L. Bajaj, M. takai, R. Ahuja, R. Bagrodia, M. Gerla, "Glomosim: a Scalable Network Simulation Environment," Technical Report 990027, Computer Science Department, UCLA, 1999.

[18] J. Klaue, B. Rathke, and A. Wolisz, "evalvid – a framework for video transmission and quality evaluation", In Proceeding of the 13th International Conference on Modeling Techniques and Tools for Computer Performance Evaluation, Urbana, Illinois, USA, 2003.

[19] http://ffmpeg.mplayerhq.hu/, ffmpeg Official Webpage, September, 2009.